\documentclass[preprint]{aastex}

\begin{document}

\title{The Tolman Surface Brightness Test for the Reality of the
Expansion. II. The Effect of the Point-Spread Function and Galaxy
Ellipticity on the Derived Photometric Parameters}

\author{Lori M. Lubin\altaffilmark{1,2}} \affil{Department of Astronomy,
California Institute of Technology,\\ Mailstop 105-24, Pasadena,
California 91125}

\author{Allan Sandage} 
\affil{Observatories of the Carnegie Institution of Washington,\\ 813
Santa Barbara Street, Pasadena, California 91101}

\altaffiltext{1}{Hubble Fellow} 
\altaffiltext{2}{Current address : Department of Physics and
Astronomy, Johns Hopkins University, Baltimore, MD 21218}

\begin{abstract}

To complete the Tolman surface brightness test on the reality of the
expansion of the Universe, we need to measure accurately the surface
brightness profiles of the high-redshift galaxy sample.  To quantify
this accuracy, we investigate the effects of various sizes of
point-spread functions composed of telescope diffraction, CCD pixel
resolutions, and ground-based seeing on the measurements of mean
surface brightness for galaxies whose effective (half-light) radii
range from 0\farcs{70} to 0\farcs{17}. We have done the calculations
using two synthetic galaxies of effective radii of 0\farcs{70} and
0\farcs{25} with point-spread functions of 0.1, 0.3, and 0.9
arcseconds, full width at half maximum. We have also compared actual
observations of three high-redshift galaxies in the cluster Cl 1324 +
3011 ($z = 0.76$) made both with the ground-based Keck 10-m telescopes
in seeing of about 0\farcs{9} and with the {\it Hubble Space
Telescope} (HST) with a point-spread function that is approximately
ten times smaller. The conclusion is that HST data can be used as far
into the galaxy image as a Petrosian metric radius of $\eta = 1.3$
magnitudes, whereas the ground-based data will have systematic errors
of up to 2.9 magnitudes in the mean surface brightness at $\eta$
values of less than 2.2 magnitudes for all but the largest galaxies at
redshifts of $z \approx 0.7$.

In the final section, we compare the differences in derived average
surface brightness for nearly circular galaxy images compared with
highly flattened images. The comparison is made by using the two
reduction procedures of (1) integrating the profile curves using
circular apertures, and (2) approximating an ``equivalent circular''
galaxy that is highly elongated by using an ``effective'' radius of
$\sqrt{ab}$, where $a$ and $b$ are the semi-major and semi-minor axis,
respectively, of the best-fitting ellipse. The conclusion is that the
two methods of reduction give nearly identical results and that either
method can be used to analyze the low and high-redshift galaxy samples
used in the Tolman test.

\end{abstract}


\section{Introduction}

More than seventy years ago, Tolman (1930, 1934) showed that, in an
expanding universe with any arbitrary geometry, the surface brightness
of a set of "standard" (identical) objects will decrease with redshift
as $(1 + z)^4$. In this series of papers, we present the necessary
steps to carry out the Tolman test using early-type galaxies which are
members of three high-redshift clusters, Cl 1324+3011 at $z = 0.76$,
Cl 1604+4304 at $z = 0.90$, and Cl 1604+4321 at $z = 0.92$ (Oke,
Postman \& Lubin 1998; Postman, Lubin \& Oke 1998, 2001; Lubin et al.\
1998, 2001). In the first paper in this series (Sandage \& Lubin 2001;
hereafter Paper I), we discussed the validity of the Tolman test,
described how we will use the observed photometric properties of
early-type galaxies to measure the surface brightness dimming as a
function of redshift, and calibrated the correlations between average
surface brightness, absolute magnitude, and linear radius for the
local galaxy sample. To use the fiducial (zero redshift) relations
derived in Paper I, we need to measure the same observable parameters
in our high-redshift galaxy sample. The purpose of this paper is to
explore possible errors in measuring these parameters from CCD imaging
and to determine how these uncertainties might affect the Tolman test.

First, we quantify the systematic errors in the determination of
Petrosian (1976) $\eta$ radii caused by a finite point-spread function
(PSF) from telescope diffraction, CCD pixel size, and the atmospheric
seeing disk if the observations are made from the ground. The
Petrosian $\eta$ radius function is defined as the difference in
magnitude between the mean surface brightness, $\langle SB (r)
\rangle$, averaged over the area interior to a particular radius and
the surface brightness, $SB(r)$, at that radius. This function is
discussed in \S 2 of Paper I where some of its unique properties in
defining a metric size are described.

While other authors have examined how seeing affects certain
photometric properties of elliptical galaxies (e.g.\ Schweizer 1979,
1981; Sparks 1988; Saglia et al.\ 1993), we need to determine
specifically the sensitivity of $\eta(r)$ to the size of the PSF. We
have done this in two ways. (1) Simulations using an assumed true
intensity profile are folded with different PSFs to give a family of
``observed'' profiles that are then used to derive the other
photometric parameters. (2) The derived $\eta$ radii for a sample of
galaxies that have been observed with both the Keck 10--m telescopes
and the {\it Hubble Space Telescope} (HST) which has a PSF about ten
times smaller are compared. The results of the simulations are given
in \S 2 for the effect on the measured photometric parameters of four
assumed PSF widths for two galaxies of different half-light angular
radii. In \S 3 we show the comparisons using actual data for three
early-type galaxies at high redshift observed both from the ground
with Keck and from space with HST by Postman et al.\ (1998, 2001) and
Lubin et al.\ (1998, 2001).

Second, we show in \S 4 the effects on $\eta(r)$ of using
elliptical-aperture photometry on highly elongated early-type galaxies
and defining the photometric parameters from the resulting intensity
profile as a function of the effective radius, $\sqrt{ab}$, where $a$
and $b$ are the semi-major and semi-minor axes, respectively, of the
best-fitting ellipse. We investigate how well the integration using
circular symmetry of an $\sqrt{ab}$ profile agrees with the
pixel-by-pixel summing in circular apertures of the actual
two-dimensional observed surface brightness array of the same
galaxies.

\section{The Effect of Various Point Spread Functions on the Measured 
Photometry Parameters by Simulations}

To assess the effects of finite point-spread-functions on measurements
of (1) mean surface brightness, (2) total magnitude, and (3) metric
radius at particular Petrosian $\eta$ values, we have simulated
observations of two elliptical galaxies with seeing and point-spread
functions whose total widths at half maximum vary from zero to
0\farcs{9}.  We have chosen the size of the intrinsic galaxy profiles
for the simulations to match closely two galaxies from the cluster Cl
1324 + 3011 at $z = 0.76$ that are included in the sample of
high-redshift, early-type galaxies in Lubin \& Sandage (2001a;
hereafter Paper III).

For the first galaxy we choose HST \#9 in Cl 1324 + 3011 which is
classified as an elliptical and which is the first-ranked cluster
member both in apparent magnitude and in angular size. The total $I$
magnitude on the Cape/Cousins photometric system is $I = 19.85$. The
effective (half-light) radius is 0\farcs{70}, as determined from the
HST image (Lubin et al.\ 2001). This is 7 pixels on the pixel scale of
0\farcs{0996} per pixel for the Wide Field Planetary Camera 2 (WFPC2)
chips.  The second galaxy for the simulation is the much fainter and
smaller elliptical cluster member, HST \#59, whose total magnitude is
$I = 22.09$ and whose effective radius is 0\farcs{25}.  For the
simulation we model the surface brightness profile of each galaxy by a
de Vaucouleurs $r^{1\over{4}}$ law as

\begin{equation}
           I = I_o~{\rm exp}~[-7.67~(r/r_{eff})^{1\over{4}}].
\end{equation}

\noindent For both simulated galaxies we assume a circular image and
perform the necessary integrations for the mean surface brightness,
growth curve, and Petrosian $\eta$ radii using circular-aperture
integration over the synthetic galaxy images.

\subsection{The Galaxy Simulations}

The artificial galaxies were created using the IRAF task MKOBJECTS.
This program allows us to create the galaxy profile, convolve it with
a seeing disk and/or PSF, add the sky background, and add Poisson
noise. For these simulations, we have first chosen a simple
representation of the seeing and have modeled the PSF as a Gaussian;
however, we also explore the effect of the more complicated PSF of the
WFPC2 on the HST observations (see \S 2.2).

For the intrinsic case of no seeing and zero width to the PSF, we have
simulated the galaxy profiles with a pixel scale of 0\farcs{010} per
pixel. This resolution is ten times better than that assumed for the
WFPC2 in the finite seeing/PSF simulations. For the calculation of the
intrinsic profile, a calculation grid as defined by a pixel scale, no
matter how small, is necessary. However, we have made the pixel scale
small enough so that the width of the intrinsic profile is effectively
zero to an excellent approximation.

For the cases where we include seeing and the PSF at various levels,
we simulate the 32 ksec HST observation of Cl 1324 + 3011 (Lubin et
al.\ 2001) by using the appropriate pixel scale of 0\farcs{0996} per
pixel and including the same background level (272 DN) of the WFPC2
observations taken in the F814W filter, which closely resembles the
$I$ band. In addition, we have added Poisson noise using the gain and
read-noise parameters of the WFPC2 which are 7 e$^{-}$/DN and 5
e$^{-1}$, respectively. To add the effect of seeing, we have convolved
the galaxy profiles with a Gaussian profile with major axis,
half-intensity diameters of 0.1, 0.3, and 0.9 arcsec, respectively.
The profiles of the two artificial galaxies, convolved by the adopted
family of PSF functions and with the noise sources added as described
above, were then used to calculate the resulting ``observed''
parameters of (1) the surface brightness profiles, (2) the growth
curves for the total magnitude, and (3) the Petrosian $\eta$ values at
various angular radii. These functions are calculated using the
profile-fitting IRAF task ELLIPSE, assuming circular aperture
integrations.
     
Curves of these functions are given in Figures 1 and 2 for HST \#9 and
\#59, respectively. The solid (red) curves in each panel represent the
``no seeing'' case. We adopt these as the intrinsic profiles for the
two galaxies. The ``observed'' curves for the three parameters of
profile, growth curve, and $\eta(r)$ are shown for the three finite
seeing PSFs.  The key for these curves are marked in the panel for the
$\eta$ function. The PSF of 0\farcs{1} (full width at half maximum)
resembles the level that can be obtained with HST, while the seeing of
0\farcs{9} is typical of observations using the Low Resolution Imaging
Spectrograph (LRIS; Oke et al.\ 1995) on the Keck 10-m telescopes.

The conclusion that will be important for the Tolman test in Lubin \&
Sandage 2001b (hereafter Paper IV) concerns the accuracy with which
the $\eta$ radii can be determined with HST given its PSF of about
0\farcs{1}. For the larger of the two simulated galaxies HST \#9, the
intrinsic $\eta$ profile is reached (to within 5\%) at $\eta = 1.1$
where log $r = -0.40$ (or $r = 0\farcs{40}$) for the 0\farcs{1} seeing
case. The intrinsic $\eta$ of the 0\farcs{3} seeing case for this
galaxy is not reached until $\eta = 1.6$ at log $r = 0.04$ (or $r =
1\farcs{10}$). For the case of 0\farcs{9} seeing (equivalent to the
Keck observations), the intrinsic $\eta$ is not reached until $\eta =
2.2$, at which place the angular radius is 2\farcs{39}. The situation,
of course, is more severe for the smaller galaxy, HST \#59. For the
0\farcs{1} seeing case, the intrinsic $\eta$ is reached (to within
5\%) only at $\eta = 1.6$ at which point the angular radius is
0\farcs{40}. For the 0\farcs{9} (ground-based) seeing case, the true
$\eta$ curve is reached only at large angular radii where $\eta >
2.5$.

The systematic error on the surface brightness averaged over the
$\eta$ radii, i.e.\ $\langle SB \rangle$, is seen by reading the
growth curves at these radii. For the case of the 0\farcs{1} PSF, the
intrinsic growth curve of HST \#9 has been reached to within the 0.02
magnitude level by log $r = -0.40$ (or $\eta = 1.1$). Hence, because
the observed $\eta$ values for the 0\farcs{1} seeing have also been
reached by $\eta = 1.1$ as discussed above, the $\langle SB \rangle$
values for galaxies which are the size of HST \#9 (half-light radius
of 0\farcs{70}) will have no systematic errors in the $\langle SB
\rangle - \eta(r)$ correlations for all $\eta$ values greater than 1.1
mag. For the smallest galaxies in Cl 1324 + 3011, such as HST \#59
whose half-light radius is 0\farcs{25}, the systematic errors in
$\langle SB \rangle$ due to the PSF of 0\farcs{1} will become
negligible only for $\eta$ values larger than 1.6 mag. This conclusion
is seen in the same way as above by noting from Figure 2 that both the
growth curve and the $\eta$ curve have reached the intrinsic curves
only by this $\eta$ value.

\subsection{The WFPC2 Point Spread Function}

The simulations presented above provide us with a general picture of
how seeing between 0\farcs{1} and 0\farcs{9} affects the galaxy
profile. Unfortunately, instrument PSFs, in particular that of WFPC2,
are considerably more complex. Although a Gaussian function represents
well the seeing disk and the initial turndown of the PSF, the wings of
the PSF typically have a shallower fall-off than a Gaussian, with a
slope that is more similar to an inverse-square power law (e.g.\ King
1971; Kurst 1995).

To explore how the exact WFPC2 PSF will affect our results, we have
used the program Tiny Tim by J.\ Kirst and R.\ Hook (see {\tt
www.stsci.edu/software/timytim}) to generate a representative PSF on
the WFC chip 3 and in the F814W filter. We then convolve this PSF with
the theoretical profiles of HST \#9 and 59 (as described above) using
the MKOBJECTS program. The results of these simulations are
qualitatively similar to those of the Gaussian seeing; however, the
larger extent of the actual PSF means that the HST profile reaches the
intrinsic (no seeing) profile at larger radii. Specifically, the
intrinsic $\eta$ curve is reached (to within 5\%) at $\eta = 1.3$
where the angular radius $r = 0\farcs{7}$ for the larger galaxy HST
\#9 and at $\eta = 1.8$ where $r = 0\farcs{6}$ for the smaller galaxy
HST \#59. At these respective $\eta$ values, the error in the measured
mean surface brightness $\langle SB \rangle$ is less than 0.07
mag. Consequently, there will be no systematic errors in $\langle SB
\rangle$ for these $\eta$ values or larger.

We will test these predictions in Paper IV for the three high-redshift
clusters studied there by noting systematic differences in the Tolman
signal that is determined from the mean surface brightness versus
metric radius correlations using $\eta$ values that range from 1.3 to
2.0 magnitudes.

\section{Comparison of Photometric Parameters of Three Galaxies Observed
With Both HST and the Keck 10-m Telescopes}

In the three clusters to be used for the Tolman test in Paper IV, all
galaxies with data from HST also have photometric data taken with LRIS
on the Keck 10-m telescopes (Oke, Postman \& Lubin 1998; Postman,
Lubin \& Oke 1998, 2001; Lubin et al.\ 1998, 2001). In this section we
compare the actual data obtained using HST with the ground-based data
for three galaxies from the cluster Cl 1324 + 3011 at a redshift of $z
= 0.76$.

Table 1 gives the characteristics of the three galaxies. The effective
radii, as measured from the HST images, range from 0\farcs{35} to
0\farcs{17}, all smaller than the effective radius for HST \#9 with
$r_{eff} = 0\farcs{70}$ described in \S 2. The smallest of the three
galaxies, HST \#69 with $r_{eff} = 0\farcs{17}$, is smaller than HST
\#59 with $r_{eff} = 0\farcs{25}$ which is also discussed in \S 2. The
apparent magnitudes in $I$ for the three galaxies range from 20.88 mag
to 22.20 mag within a circular aperture of $3''$ in radius.  These
values closely match the asymptotic total magnitudes of the growth
curves in Figures 3, 4, and 5.  Both the HST and the Keck images of
the three galaxies have been reduced in a manner similar to that
described in \S 2. Here, however, we have fit elliptical apertures to
the two-dimensional data so as to get the most accurate representation
of the galaxy profile (see also \S 4).  The resulting curves of
surface brightness, total magnitude, and Petrosian radii $\eta$, which
are shown in Figures 3 -- 5, are plotted against log $a$, where $a$ is
the semi-major axis of the best-fitting elliptical aperture. The
curves are derived from data taken in the Cape/Cousins $I$ band for
the Keck observations and the F814W filter for the HST
observations. The two bandpasses are very similar with $F814W = I +
0.05$.

Figure 3 shows the comparison for HST \#18. It is the largest of the
three galaxies and has photometric characteristics which are
approximately the mean of the two galaxies presented in \S 2 and shown
in Figures 1 and 2. The Keck $\eta$ curve does not reach the
equivalent HST $\eta$ curve until $\eta \approx 2.2$.  Therefore, the
measured angular radii from the Keck data are too large compared to
the true (HST) $\eta$ value by factors that range from a factor of 2.2
at $\eta = 1$, decreasing to a factor of 1.7 at $\eta = 1.5$. Although
the difference in the total magnitude between the HST and Keck
profiles decreases with increasing radius, it is still 0.15 magnitudes
too faint at a radius of $2''$. Hence, for galaxies with $r_{eff} =
0\farcs{35}$, no reliable mean surface brightness values can be
obtained from ground-based data with the LRIS seeing level even at
$\eta$ values of 2.  That is, the inferred $\langle SB \rangle$ values
from the Keck ground-based data will be too faint by 1.7 mag at $\eta
= 1$ and 1.3 mag at $\eta = 1.5$.

The situation becomes progressively worse, of course, as the effective
radius decreases. Figure 4 shows the case for HST \#40 with $r_{eff} =
0\farcs{26}$. Here, the HST $\eta$ radius is never attained by the
measured Keck $\eta(r)$ curve for any reasonable values of $\eta$.
Figure 5 shows the case for $r_{eff} = 0\farcs{17}$ where the errors
due to seeing are the worst of the three. At a fixed $\eta$ value, the
mean surface brightness is underestimated by up to 2.9 magnitudes when
the Keck data are used to measure the profiles of these galaxies. The
clear conclusion from these comparisons is that, for small galaxies,
no reliable data on surface brightness can be obtained from
ground-based observations with seeing of 0\farcs{9}.  Even for the
largest galaxies at $z \approx 0.7$, only profile data at Petrosian
radii values of $\eta > 2$ can be accurately used.  This conclusion
is, of course, not a surprise; however, it is here made quantitative.

It should also be noted that deconvolution methods can be used to
recover the original galaxy profile if the PSF of the instrument can
be accurately measured. Such methods have improved considerably over
the past years due to detailed analyses of images from WFPC and WFPC2
aboard the {\it Hubble Space Telescope}. For this project, we have
chosen not to employ any deconvolution methods, but rather to use
profile data only where we can be assured of the measurement accuracy.
We have done this because the point-spread-functions of both the WFPC2
and the LRIS CCDs are a strong function of chip position. For example,
the ellipticity of the PSF varys from 1\% at the chip center to
15--25\% at the chip edge (see e.g.\ Krist 1995; Hoekstra et al.\
1998; Clowe et al.\ 2000; Squires et al.\ 2001). In addition, the PSF
of the LRIS CCD is dependent on the rotator position. Consequently,
the PSF as a function of chip position has to be measured separately
for each LRIS image (Squires et al.\ 2001).  Because we are analyzing
galaxy profiles across the entire field for both the HST and Keck
observations, we would need to measure the PSF at all chip positions.
Since we have shown in \S 2 that the high-angular-resolution HST data
is a reasonable measure of the true galaxy profile over the range of
Petrosian radii in which we are interested ($\eta \gtrsim 1.3$), we
have chosen to use the HST profiles without deconvolution.

\section{Comparison of Photometric Parameters Measured With Circular Apertures
on Highly Elongated Galaxies}

\subsection{Various Procedures for Reducing Photometric Data for Highly 
Elongated Early-Type Galaxies}

A persistent problem in the photometry of galaxies is how to treat the
photometric data to obtain Petrosian radii, mean surface brightness,
and magnitudes inside given apertures for early-type galaxies that are
highly elongated.  In the methods of single aperture photoelectric
photometry done between 1950 and 1980 before two dimensional array
detectors were available, the focal plane blocking apertures at the
telescope in the photometers were circular. The growth curves were
obtained directly from the data as the sizes of the blocking apertures
were increased. All of the aperture photometry and the growth curves
in the literature of the 1970's on the velocity-distance relation
(e.g.\ Sandage 1972a,b,c; 1973) were made in this way.

When two-dimensional photoelectric data became available with the
invention of areal detectors, a number of different decisions had to
be made on how to present and to analyze the data. Among the many
methods used in the presentation of the data are (1) to give the
intensity profile (mag per arcsec$^2$ at a particular radii) along the
major axis, listing the {\it semi-major axis} angular distances from
the center, (2) the same profile as in (1) but listed using the
``effective radius'' defined as $\sqrt{ab}$ for the best-fitting
ellipse, or (3) an ``effective profile'' defined as the projection of
the two-dimensional data onto the major axis and listed by semi-major
axis angular distances (see Watanabe, Kodaira \& Okamura 1982).

The growth curves of magnitudes versus radii can be obtained either by
integrating the one-dimensional profile or by using the
two-dimensional data in several ways. First, numerical aperture
photometry can be done by summing the intensities, pixel-by-pixel, in
circular areas of growing size. This imitates the 1970's growth curve
observations made at the telescope with single--channel photometers
using circular blocking apertures.  Second, a more elaborate method is
to fit elliptical isophotal contours to the two-dimensional data and
to numerically sum the intensities using elliptical apertures. This
procedure is equivalent to observations using a single channel
photometer at a telescope by using different sized blocking diaphragms
that would be elliptical to fit the galaxy contours.  The resulting
growth curve, plotted as a function of some measure of the angular
radii, either the semi-major axis distance, $a$, or the ``effective''
radii, $\sqrt{ab}$ (Lauer 1985). Third, an integration can be made of
the profile data using any of the modes of presentation enumerated
earlier using a circular-aperture assumption for the integration over
an area.  Generally, the results for the Petrosian metric radii
$\eta(r)$, the average surface brightness $\langle SB \rangle$, and
the growth curve will differ systematically among the methods.

All of these methods and more have been used in the literature,
although sometimes there is insufficient information to tell which
profile and which definition of angular radius is used.  The purpose
of this section is to make explicit the methods that we have used for
Papers III and IV, as well as to compare these methods with those that
are used in Paper I to make the local calibrations. Specifically, the
data in Paper I were obtained from circular-aperture photometry on all
of the local galaxies regardless of their ellipticity (see \S 3 of
Paper I and below).
          
\subsection{Comparison of Circular-Aperture Photometry With $\sqrt{ab}$ 
Profile Photometry for Highly Elongated Galaxies}

Postman \& Lauer (1995) reduced their two-dimensional CCD photometry
of the first ranked cluster galaxies in their sample of 119 Abell
clusters by using circular-aperture integrations, pixel-by-pixel, from
the two-dimensional image arrays. Their circular-aperture growth-curve
magnitudes for various radii are listed in their Table 3.  We have
used these data in Paper I to define the zero redshift calibrations of
the $\langle SB \rangle$, Petrosian radii $\eta(r)$, and growth-curve
correlations.  On the other hand, the correlations for the local
calibrations used by Sandage \& Perelmuter (1991a,b) to derive similar
calibrations using independent data from the literature sources cited
in Paper I were made by integrating profile data that were generally
given as a function of the ``effective'' radius, $\sqrt{ab}$. That is,
the surface brightnesses were measured along the semi-major axis, but
the resulting surface brightness profile was listed as a function of
radius $r = \sqrt{ab}$.  This profile is often referred to as the
``equivalent circular'' galaxy that, to some degree, imitates a highly
flattened galaxy.  But how close is the correspondence?

To explore this correspondence, we compare in this section three
different methods for analyzing the two-dimensional CCD images and
calculate the resulting curves of surface brightness, total magnitude,
and $\eta$. The three methods are as follows :

\newcounter{discnt}
 
\begin{list}
{\arabic{discnt}.}  {\usecounter{discnt}}

\item 
The two-dimensional data for the galaxy image are analyzed by using a
series of circular apertures, regardless of the ellipticity of the
galaxy. The surface brightness profile (in mag per arcsec$^2$) is
measured at the circular radius $r$, and the growth curve is derived
by summing the pixel intensities within each circular aperture.

\item The two-dimensional data for the galaxy image are fitted by a series
of ellipses of progressively larger semi-major axis sizes. The surface
brightnesses are measured along the major axis, and the growth curve
is derived by summing the pixel intensities within each elliptical
contour.  The resulting parameters are listed against the semi-major
axis $a$.

\item  The tabulated surface brightness (SB) profile as measured in item
(2) is used; however, it is plotted as a function of $r = \sqrt{ab}$
for the radii, instead of the semi-major axis, $a$.  This profile is
then re-integrated under a circular-aperture assumption as
$\int{SB~2\pi r dr}$ in order to obtain the mean surface brightness,
growth curve, and $\eta(r)$.  This is said to approximate the same
parameters of an ``equivalent circular'' galaxy.

\end{list}

In Papers III and IV, we shall use the $\sqrt{ab}$ radii analysis
(item 3) for the high-redshift cluster galaxies. However, the data
which was used in Paper I for the fiducial (local) lines was reduced
by Postman \& Lauer (1995) with circular apertures placed on the
observed two-dimensional array (item 1). For our analyses in Paper I,
we did not need the profile data as the growth curves were all that
were necessary to derive $\langle SB(\eta) \rangle$ and $\eta(r)$.

Since we are using two different data analysis methods for the Tolman
test, it is essential to assess how well the $r = \sqrt{ab}$
``equivalent circular galaxy'' assumption used in Papers III and IV
agrees with the circular aperture reductions for highly elongated
galaxies presented in Paper I. We have made the comparison using two
galaxies in the cluster Cl 1324 + 3011 at a redshift of $z =
0.76$. The first galaxy HST \#11 (Lubin et al.\ 2001) is nearly
circular with an minor-to-major axis ratio of $b/a = 0.9$ near the
center, flattening to $b/a = 0.75$ far from the center. The comparison
galaxy is HST \#12 which is highly elongated with an average
minor-to-major axis ratio of $b/a = 0.50$. The detailed data on these
galaxies, which are based on original data from Lubin et al.\ (2001),
are set out in Paper III where all of the photometry for the
Tolman-test clusters is listed.

Figures 6 and 7 show the digital images of HST \#11 and HST \#12 from
the WFPC2 data which was taken in the F814W filter.  Figures 8 and 9
show the diagnostic diagrams for HST \#11 and \#12 determined from the
three methods described above. The solid blue circles show the results
of the circular aperture analysis (item 1); the solid cyan squares
represent the elliptical aperture analysis where the resulting curves
are plotted against semi-major axis $a$ (item 2); and the open red
circles represent a re-analysis of the elliptical aperture data using
the effective circular radius of $\sqrt{ab}$ (item 3).

We first discuss Figure 8 which presents the results for the nearly
circular galaxy of HST \#11. As expected for a nearly round galaxy,
all three curves in each panel give very consistent results for all
radii except those in the very center (i.e.\ log $r \lesssim -0.5$ or
$\eta \lesssim 1$).  In order to carry out the Tolman test, we must
compare the low and high-redshift data using both the mean surface
brightness and the linear radius at a given $\eta$.  Therefore, the
two most crucial diagrams are $\langle SB \rangle$ versus $\eta$
(lower right panel) and $\eta$ versus radius (upper right panel). The
conclusion is that for values of $\eta > 1$ all three methods give
nearly identical results for the mean surface brightness at a given
$\eta$. The average difference between all three methods is $\Delta
\langle SB \rangle = 0.07$ magnitudes for all Petrosian radii with
$\eta > 1$.

We note, however, that in the $\eta$ -- log $r$ panel, even for the
nearly circular galaxy HST \#11, the curve derived using elliptical
apertures (solid cyan squares) begins to deviate systematically from
the other two curves at large $\eta$ where the ellipticity is highest
(i.e.\ $b/a = 0.75$). At $\eta > 1.4$, the average difference between
the circular aperture and elliptical apertures method is $\Delta({\rm
log}~r) = 0.10$ (a 26\% error in $r$). However, this difference is
only $\Delta({\rm log}~r) = 0.03$ (a 7\% error in $r$) between the
circular aperture and $\sqrt{ab}$ methods. This is, of course, because
the semi-major axis $a$ is larger than the ``equivalent circular''
galaxy using the $\sqrt{ab}$ approximation. Hence, even though the
lower right panel shows near agreement in the $\langle SB \rangle -
\eta$ plane at mid-$\eta$ values, the systematic differences in the
$\eta$ -- log $r$ plane will cause a systematic difference in any
comparison if one set of data would be reduced using the elliptical
aperture (major axis) method, while another set would be reduced by
either the circular aperture method or the closely equivalent
$\sqrt{ab}$ method.

These differences are, of course, larger for the highly elongated
galaxy HST \#12 in Figure 9. Here, the increased radius of the major
axis method is well seen in the upper left panel of surface brightness
($SB$) versus log (radius). However, as in Figure 8, the $\langle SB
\rangle$ versus $\eta$ panel at the lower right shows only a modest
spread between the three methods for mid-$\eta$ values. The average
difference is $\Delta \langle SB \rangle = 0.13$ magnitudes for all
Petrosian radii of $1 < \eta \le 2$. Again, the important panel to
consider is the $\eta$ -- log $r$ panel at the upper right. While the
elliptical aperture method is substantially different, the circular
aperture (solid blue circles) and the $\sqrt{ab}$ (open red circles)
methods define nearly the same $\eta$ curve to within 0.03 in log $r$
for all radii with log $r > -0.4$. This proves again that use of the
$\sqrt{ab}$ measure of the ``effective'' radius is closely equivalent
to the use of circular-aperture photometry. We make use of this
important conclusion in Papers III and IV in the discussion of the
high-redshift cluster photometry.

\section{Conclusions}

In order to perform successfully the Tolman test using data from
high-redshift cluster galaxies, we must make a reliable measure of the
galaxy surface brightness profiles which are used to derive the
essential parameters of mean surface brightness, total magnitude, and
the Petrosian metric radius. In this paper, we have quantified the
reliability of these quantities, first, by measuring the effect of
seeing on the galaxy profiles and, second, by comparing the various
methods of reducing the galaxy imaging data. Using both simulations of
galaxy profiles convolved with Gaussian seeing from zero to 0\farcs{9}
and actual data obtained with the {\it Hubble Space Telescope} and the
Keck telescopes, we have found that high-angular-resolution data
similar to that of HST can be used reliably to measure galaxy
properties as far into the galaxy image as a Petrosian metric radius
as small as $\eta \approx 1.3 - 1.8$, depending on the angular size of
the galaxy.  On the contrary, ground-based data taken in poor seeing
(full width at half maximum of 0\farcs{9}) have systematic errors of
up to 2.9 magnitudes in the mean surface brightness for all Petrosian
metric radii with $\eta \lesssim 2.2$ for all but the largest galaxies
at redshifts of $z \approx 0.7$.

While the imaging data on the local galaxies which will serve as our
zero-redshift calibration for the Tolman test were originally analyzed
using circular apertures, we have analyzed the CCD images of our
high-redshift galaxy sample through the ``equivalent circular galaxy''
method which gives the surface brightness profile (measured in
elliptical apertures) as a function of the effective radius,
$\sqrt{ab}$. We have shown that both methods are consistent to within
7\% for all Petrosian radii values of $\eta > 1$ for nearly circular
galaxies, as well as highly elongated galaxies. Hence, the Tolman test
is practical using these methods because an accurate comparison
between the profile data of the low and high-redshift galaxy samples
can be made.

\acknowledgments 

We thank the anonymous referee for suggesting useful additions to this
paper. We also thank Marc Postman and J.B.\ Oke for their permission
to use part of their extensive Keck and HST database in this paper.
LML is supported by NASA through Hubble Fellowship grant
HF-01095.01-97A from the Space Telescope Science Institute, which is
operated by the Association of Universities for Research in Astronomy,
Inc., under NASA contract NAS 5-26555. AS acknowledges support for
publication from NASA grants GO-5427.01-93A and GO-06549.01-95A for
work that is related to data from HST.

\newpage

\figcaption[]{The effect of various point spread functions and/or
seeing disks on the photometric parameters for a galaxy with an
effective (half-light) radius of 0\farcs{70}. The curves in each panel
are derived from a synthetic galaxy which has a de Vaucouleurs
$r^{1\over{4}}$ profile and is circularly symmetric. The three panels
show the surface brightness profile (upper left), the Petrosian $\eta$
function (upper right), and the growth curve of the total magnitude
within a given radius $r$ (lower left), all as a function of log $r$
where $r$ is in arcsec. The normalizations of the ordinates for the
surface brightness and total magnitude profile are from the actual
data for the galaxy HST \#9 in the cluster Cl 1324+3011 ($z = 0.76$)
as given in Paper III.  The solid red line in each panel indicate the
intrinsic relation for zero seeing, while the remaining three lines
indicate the relations with Gaussian seeing of 0\farcs{1} (blue dots),
0\farcs{3} (magenta short dashes), and 0\farcs{9} (cyan long dashes),
respectively.}

\figcaption[]{Same three panels as Figure 1, but for a simulated
galaxy of half-light radius 0\farcs{25} which closely resembles galaxy
HST \#59 in Cl 1324+3011.}

\figcaption[]{Same panels as given in Figures 1 and 2, but using
actual observations of the galaxy HST \#18 in Cl 1324+3011 made with
both the {\it Hubble Space Telescope} (solid circles) and the Keck
10-m ground-based telescopes (open circles). The bandpass is the
Cape/Cousins $I$ band for the Keck data and the F814W filter for the
HST data, respectively (see \S 3).}

\figcaption[]{Same panels as Figure 3, but using actual observations
of the galaxy HST \#40 in Cl 1324+3011 made with both HST and the Keck
telescopes.}

\figcaption[]{Same panels as in Figures 3 and 4, but for the galaxy
HST \#69 in Cl 1324+3011 that was observed both with HST and the Keck
telescopes. This galaxy is the smallest of the three galaxies which
are presented in Figures 3--5.}

\figcaption[]{CCD image of the galaxy HST \#11 in the cluster Cl
1324+3011 taken from the full WFPC2 image of a 32.0 ksec observation
in the F814W filter. The image is $5'' \times 5''$.}

\figcaption[]{Same as Figure 6, but for the highly elongated
elliptical galaxy HST \#12 in Cl 1324+3011.}

\figcaption[]{The difference in the photometric correlations using the
three methods discussed in \S 4 for reducing the two-dimensional
intensity data.  The four panels show the surface brightness (upper
left), Petrosian $\eta$ function (upper right), and growth curve
(lower left) as a function of radius (either circular radius,
semi-major axis, or $\sqrt{ab}$) in arcsec, and the mean surface
brightness as a function of $\eta$ (lower right). The curves are
derived from data taken with HST for the galaxy HST \#11 in Cl
1324+3011. The solid blue circles, solid cyan squares, and open red
circles indicate the circular aperture, elliptical aperture, and
$\sqrt{ab}$ methods, respectively (see \S 4.2).}

\figcaption[]{Same four panels as Figure 8, but for the highly elongated
elliptical galaxy HST \#12 in Cl 1324+3011.}

\newpage

\begin{deluxetable}{ccccccccc}
\tablewidth{0pt}
\tablenum{1}
\tablecaption{Properties of the Galaxies from Cl 1324+3011 used in the HST versus Keck Comparison}
\tablehead{
\colhead{HST \#\tablenotemark{a}} &
\colhead{Class\tablenotemark{a}} &
\colhead{$m_{814}$\tablenotemark{a}} &
 \colhead{$r_{eff}$\tablenotemark{a}} &
\colhead{$B$\tablenotemark{b}} &
\colhead{$V$\tablenotemark{b}} &
\colhead{$R$\tablenotemark{b}} &
\colhead{$I$\tablenotemark{b}} &
\colhead{$z$}}
\startdata
18	& E/S0	& 20.85	& 0.35	& 25.25   &  23.38 &  22.14 &  20.88 & 0.7592 \\
40	& E	& 21.69	& 0.26	& 25.91   &  23.98 &  23.22 &  21.60 & 0.7618 \\
69	& S0	& 22.26	& 0.17	& \nodata &  24.29 &  23.89 &  22.20 & 0.7628 \\
\enddata

\tablenotetext{a}{The identification number, visual morphological
classification, total galaxy magnitude, and effective radius (in
arcsec) of the galaxy surface brightness profile as measured from the
WFPC2 image taken in the F814W filter (see Lubin et al.\ 2001).}

\tablenotetext{b}{Keck $BVRI$ photometry computed in a circular
aperture with radius of $3''$ (see Postman, Lubin \& Oke 2001).}

\end{deluxetable}


\newpage

\begin{thebibliography}{}

\bibitem[]{} Clowe, D., Luppino, G.A., Kaiser, N., \& Gioia, I.M.\
2000, ApJ, 539, 540
\bibitem[]{} Hoekstra, H., Franx, M., Kuijken, K., \& Squires, G.\
1998, ApJ, 504, 636
\bibitem[]{} King, I.R.\ 1971, PASP, 83, 199 
\bibitem[]{} Krist, J.E.\ in Calibrating Hubble Space Telescope, Post
Servicing Mission; eds.\ A. Koratkar \& C.\ Leitherer; Proceedings of
a Workshop held at Space Telescope Science Institute, Baltimore 1995
(STScI; Baltimore)
\bibitem[]{} Lauer, T.\ 1985, ApJS, 57, 473
\bibitem[]{} Lubin, L.M., Postman, M., Oke, J.B., Ratnatunga, K.U.,
Gunn, J.E.,  Hoessel, J.G., \& Schneider, D.P.\ 1998, AJ, 116, 584
\bibitem[]{} Lubin, L.M., Postman, M., Oke, J.B., Brunner, R., Gunn,
J.E., \& Schneider, D.P.\ 2001, AJ, in preparation
\bibitem[]{} Lubin, L.M., \& Sandage, A.\ 2001a, AJ, in preparation
(Paper III)
\bibitem[]{} Lubin, L.M., \& Sandage, A.\ 2001b, AJ, in preparation
(Paper IV)
\bibitem[]{} Oke et al.\ 1995, PASP, 107, 375 
\bibitem[]{} Oke, J.B., Postman, M., \& Lubin, L.M.\ 1998, AJ, 116, 549
\bibitem[]{} Petrosian, V.\ 1976, ApJ, 209, L1
\bibitem[]{} Postman, M. \& Lauer, T.\ 1995, ApJ, 440, 28 
\bibitem[]{} Postman, M., Lubin, L.M., \& Oke, J.B.\ 1998, AJ, 116, 560
\bibitem[]{} Postman, M., Lubin, L.M., \& Oke, J.B.\ 2001, AJ, in preparation
\bibitem[]{} Saglia, R.P., Bertschinger, E., Baggley, G., Burstein, D., 
Colless, M., Davies, R.L., McMahan, R.K., \& Wegner, G.\ 1993, MNRAS, 264, 961
\bibitem[]{} Sandage, A.\ 1972a, ApJ, 173, 485 
\bibitem[]{} Sandage, A.\ 1972b, ApJ, 178, 1
\bibitem[]{} Sandage, A.\ 1972c, ApJ, 178, 25
\bibitem[]{} Sandage, A.\ 1973, ApJ, 183, 711
\bibitem[]{} Sandage, A., \& Lubin, L.M.\ 2001, AJ, submitted (Paper I) 
\bibitem[]{} Sandage, A., \& Perelmuter, J-M.\ 1990, ApJ, 361, 1 
\bibitem[]{} Sandage, A., \& Perelmuter, J-M.\ 1991, ApJ, 370, 455
\bibitem[]{} Schweizer, F.\ 1979, ApJ, 233, 23
\bibitem[]{} Schweizer, F.\ 1981, AJ, 86, 662
\bibitem[]{} Sparks, W.B.\ 1988, AJ, 95, 1569
\bibitem[]{} Squires, G.K.\ et al.\ 2001, ApJ, in preparation
\bibitem[]{} Watanabe, M., Kodaira, K., \& Okamura, S.\ 1982, ApJS, 50, 1

\end{thebibliography}
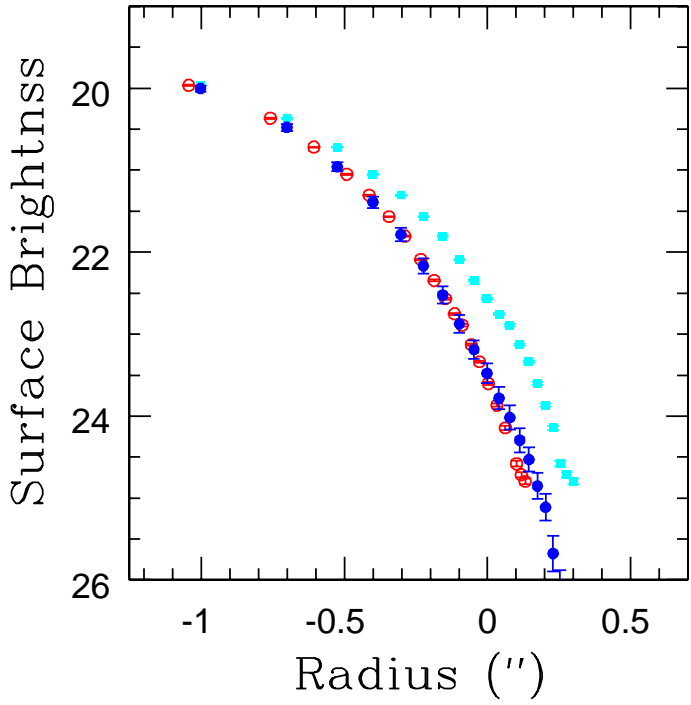
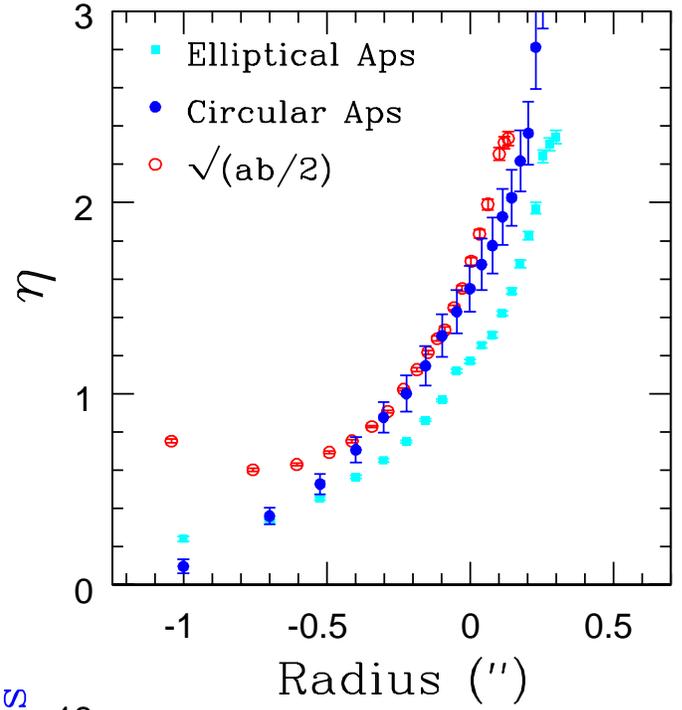
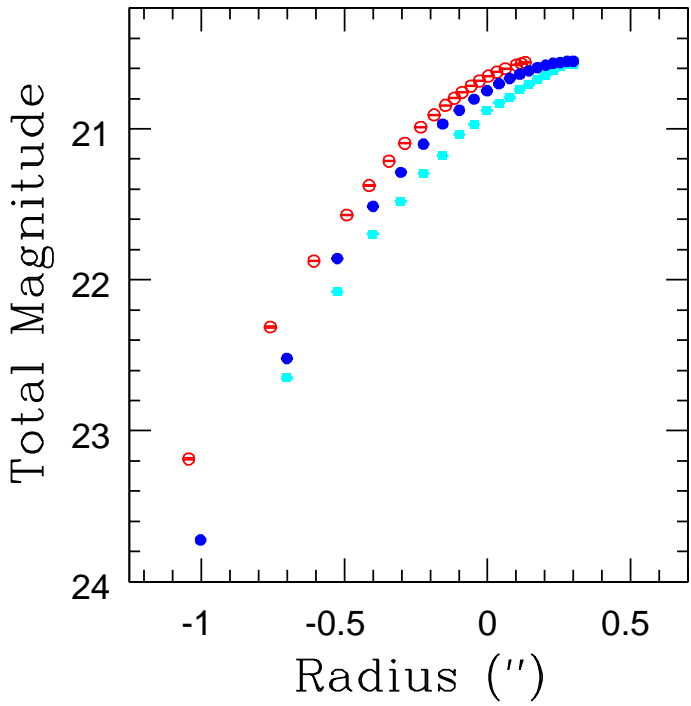
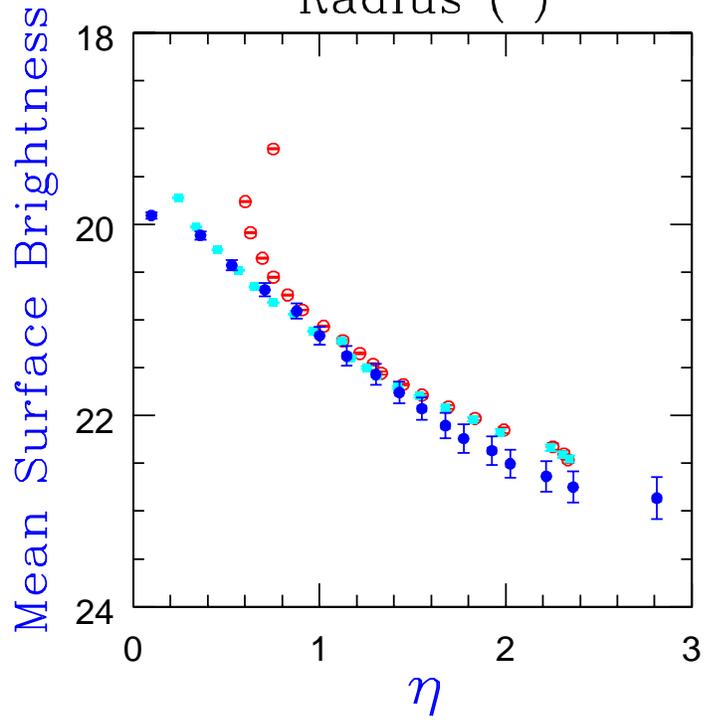
\end{document}